# Smart Traction Control Systems for Electric Vehicles Using Acoustic Road-type Estimation

Daghan Dogan, Pinar Boyraz*, *Member, IEEE*

***Abstract*— The application of traction control systems (TCS) for electric vehicles (EV) has great potential due to easy implementation of torque control with direct-drive motors. However, the control system usually requires road-tire friction and slip-ratio values, which must be estimated. While it is not possible to obtain the first one directly, the estimation of latter value requires accurate measurements of chassis and wheel velocity. In addition, existing TCS structures are often designed without considering the robustness and energy efficiency of torque control. In this work, both problems are addressed with a smart TCS design having an integrated acoustic road-type estimation (ARTE) unit. This unit enables the road-type recognition and this information is used to retrieve the correct look-up table between friction coefficient and slip-ratio. The estimation of the friction coefficient helps the system to update the necessary input torque. The ARTE unit utilizes machine learning, mapping the acoustic feature inputs to road-type as output. In this study, three existing TCS for EVs are examined with and without the integrated ARTE unit. The results show significant performance improvement with ARTE, reducing the slip ratio by 75% while saving energy via reduction of applied torque and increasing the robustness of the TCS.**

NOMENCLATURE

| | | | |
|---|---|---|---|
| $\lambda$ | Slip ratio | $\tau_2$ | Time constant for high pass filter in MFC |
| $I_{com}$ | Current command | $A$ | Gradient of μ/λ curve |
| $w$ | Rotational speed of wheel shaft | $M$ | Vehicle mass |
| $J$ | Moment of inertia for vehicle body | $I$ | Vehicle inertia |
| $J_W$ | Moment of inertia for wheel and its shaft | $V$ | Velocity of vehicle |
| $r$ | Tire radius | $W$ | Normal force of wheel |
| $F_m$ | Motor torque | $T$ | Motor torque |
| $F_d$ | Friction force | $F_{dr}$ | Driving resistance |
| $M_w$ | Wheel inertia | $V_W$ | Wheel velocity |
| $N$ | Vertical force | $\mu$ | Friction coefficient |
| $g$ | Gravity acceleration | $\alpha$ | Relaxation factor |
| $F_r$ | Rolling resistance | $\dot{V}$ | Acceleration |
| $F_a$ | Air resistance | $\dot{w}$ | Angular acceleration |
| $\tau_m$ | Motor time constant | $m$ | Wheel mass |
| $\tau_1$ | Time constant for electric motor in MFC | $\tau_1$ | Time constant for low pass filter in MTTE |

D. D. Author is with the Istanbul Technical University and TUBITAK, Turkey. (e-mail: daghan.dogan@tubitak.gov.tr).

*P. B. Corresponding Author was with Mech. Eng. Dept. at Istanbul Technical University, Turkey. She is now with the Dept. of Mechanics and Maritime Sciences, M2, Chalmers University of Technology, Göteborg, Sweden (e-mail: pinar.boyraz@chalmers.se).

## I. INTRODUCTION

THERE are currently two focus areas in vehicle technology research; the first one is based on electric vehicles and the second one involves the active vehicle safety systems towards fully autonomous cars. The active safety systems in automated vehicles can be ranging from TCS (traction control system) and ABS (anti-lock braking system) to LKS (lane keeping system) and DYC (direct yaw control). As an intersection of electric vehicles and active safety research, direct-drive motors offer a new application field to improve the existing active vehicle safety concepts while benefiting from embedded mechatronics interfaces and advanced signal processing. This type of systems can be more feasible in EVs compared to conventional vehicles with ICE. As it was suggested in [1], a complete control system addressing longitudinal and lateral vehicle dynamics becomes possible if the necessary signal processing algorithms and controllers can be applied on EVs. One such application is TCS which can be considered as a sub-system for many of the upper-level vehicle stability control systems with the aim to minimize the slip and skid of the wheels. This is achieved by conserving the necessary adhesion between the road and the wheel. The applications of TCS as a sub-part of a more comprehensive active vehicle safety system such as DYC can be seen in literature [2,3,4].

There are several types of TCS in literature in addition to dedicated studies examining road-tyre friction characteristics focusing on temperature [5] and prolonged sliding conditions [6]. If the TCS application is narrowed down to EVs, one of the simplest form is model following controller (MFC) studied in [1]. MFC system represents the longitudinal dynamic behaviour of the vehicle body in an internal model and generates wheel reference angular velocity. Then, wheel reference angular velocity is compared with actual angular velocity value to generate an error signal. After high-pass filter conditions the error signal, it is amplified and fed back to control the torque of the electric motor in turn. MFC system is very basic and straightforward using the electric motor embedded in the wheel; therefore, it helps reaching the demanded traction forces via motor torque control. However, its major disadvantage is the lack of updated road conditions. The real dynamics between the road and tire in terms of slip ratio and friction coefficient are not measured by MFC system at all. The slip ratio used in the calculation of the real vehicle inertia is a pre-set, constant value which does not update itself dynamically according to the current condition of the road. Another essential TCS for EVs named as Slip Ratio Control (SRC) [7,8,9,10] system. In SRC, velocity difference between vehicle chassis and the wheel provides the actual slip ratio. As a reference to be compared with the measured slip ratio, optimal slip ratio is estimated using an either a fuzzy inference system or gradient descent algorithm involving μ-lambda friction characteristic curves and formulae. The signals of a driving force observer are also included in SRC with the fuzzy system, yielding indirect but essential indications reflecting the level of adhesion between tire and road. Although the estimation of road characteristics represented by optimal slip ratio is just an estimated value, the actual road-tire dynamics are at least considered in SRC. Following the actual traction force between tire and road, the force observer mainly determines the performance of SRC. A second source of uncertainty may be from the look-up tables of μ-lambda curves of road conditions (i.e. mu-lambda curves obtained by Magic Formula [11]) to get an optimal lambda ($\lambda_{opt}$) using fuzzy system. The double estimation process is used to obtain the optimum slip ratio which is then tracked by adjusting the motor torque. In brief, although estimation errors of SRC may decrease its performance, it could perform better than MFC system because of its dynamic update on the road conditions. Lastly, the Maximum Transmissible Torque Estimation (MTTE) can be mentioned as the third main TCS [12, 13] for EVs. MTTE needs neither chassis velocity nor the estimation of road-tire condition. Thus, it is fundamentally different from SRC as structure. The uncertain dynamics between the road and tire together with its effect on the chassis-wheel velocity difference is accounted in the relaxation term alpha ($\alpha$) and an observer-based strategy is followed in MTTE.

In addition to these TCS structures, it is worth mentioning several recent studies on control sub-systems or sensor platforms for traction control. For example, the coefficient of tire-road was estimated using novel wireless piezoelectric sensors, to measure tire sidewall deflections and employ the conventional tire brush model to complete the estimation in [14]. This study again emphasizes the necessity of a plausible model to obtain the road-tire friction coefficients although a specialized sensor is developed. Being relatively new, acoustic methods are also employed in road condition estimation in a limited capability distinguishing between the dry and wet asphalts in [15]. Estimation of the frictional condition is used with an adaptive vehicle speed control in [16]. To obtain an estimation of the frictional condition, they employed the actuation redundancy of an in-wheel-motor (IWM) configured EV. This method seems to identify the friction characteristics of the particular road without longitudinal motion of the vehicle, therefore it is innovative. The traction control applications can go as far as the method used in [17] controlling the wheel slip employing a sliding mode controller and having a sliding mode observer to estimate the friction value. However, the controller design in that work assumes that the velocity or angular velocity measurements are available. In [18], another application, a sliding-mode observer is proposed. Similar examples involving wheel slide protection (WSP) in railway field can be found in [19, 20]. There are also real-time applications of traction control/antiskid systems [21] usually applied in a separate rig for railway vehicles [22]. Besides the studies which are focused on the traction control, there are also numerous studies on estimation of the road-tire friction in general, independent of its end-use. An early example of slip-ratio based road friction estimation was detailed in **[23]** using the standard sensors in ABS. Later a similar approach **[24, 25, 26]** was taken for wheel slip control for ABS brakes, employing gain scheduling using LQR control. There are also studies applying recursive identification for the cornering stiffness estimation **[27]** and noise-adaptive particle filtering **[28]** to provide a better understanding of road-tire dynamics in lateral motion of the vehicle. In more comprehensive works, road frictional coefficient can be estimated using a nonlinear observer **[29]** but both longitudinal and lateral models for vehicle dynamics is needed. A very interesting low-cost application **[30]** tracks the vehicle speed using chassis vibration when GPS or wheel-based sensors are not available. Although it does not aim for road-friction estimation directly, it can be used as an alternative low-cost sub-system for the TCS which might require slip-ratio measurements. One of the recent studies uses in-tire accelerometers [31] to determine the tire deformation directly to be used in driving, braking or other stability systems. There are also efforts in the identification of road-tire friction forces in real time [32] using curve fitting algorithms. Although the previous studies contributed greatly to the estimation of the road-tire interaction characteristics and proposed several TCS, there is still need for an affordable sensor platform which can be integrated into TCS structures increasing its robustness while considering the energy efficiency of the torque control.

In this work, an acoustic signal-based road-type estimator is proposed which is capable of distinguishing between asphalt, gravel, snow and stone. The unit is called 'Acoustic Road Type Estimation' system or *ARTE* in short. Signal processing methods in acoustics and speech recognition fields are adopted for extraction of features. Then artificial neural networks (ANN) or support vector machines (SVM) is used for classification. To demonstrate the effect of ARTE unit on robustness and energy efficiency of traction control, three TCS structures are examined first without, then with the unit integrated. The resultant smart TCS with ARTE unit could reduce the slip ratios and decrease the required torque effort. Furthermore, ARTE unit provides the TCS with more robustness due to reduced uncertainty in the input. Although the main contribution is in providing the up-to-date slip-ratio for the control algorithm in an affordable way, the research also derives transfer functions for three TCS and analyses the robustness of them with and without the ARTE unit. This extra outcome of the research can help researchers working in TCS design as a guideline to increase the robustness and energy efficiency of their systems. The real innovation value of ARTE is to provide a (1) low-cost and easy to set-up alternative for road-condition estimation, (2) integration of the road-condition estimation in already existing TCS for EV.

The manuscript is organized to reflect the design of ARTE unit first and then its integration with TCS, finally concluding on the performance comparison of TCS with and without the proposed unit. Section 2 describes the ARTE unit including the data acquisition platform, signal processing, and selection of acoustic features and road-type classification part using ANN or SVM. Then, in Section 3 three TCS structures for EVs mainly based on MFC, SRC and MTTE are introduced and their transfer functions are derived to be used in robustness analysis. In Section 4, the ARTE unit and selected TCS are integrated and the performance of TCS with and without the ARTE unit is compared. Lastly, the conclusions are drawn in Section 5 together with the future improvement and expansion of the proposed smart TCS.

## II. Acoustic Road Type Estimation (ARTE) System

There are many mainstream methods available for friction force estimation or measurement. However, they often require high-cost sensors and complicated set-ups. The method described herein aims to offer a simplified and low-cost system as an alternative for road-tire friction estimation and a methodology to integrate this system in TCS structures. In order to have a clear comparison, the mainstream methods are discussed briefly. A great portion of friction estimation systems uses the vehicle dynamics, wheel motion or tire-deformation as the basis for measurement or estimation. They can be considered in three main groups as follows:

(I) *Vehicle-based friction estimation system model:* uses vehicle's longitudinal and lateral motion measurement for the estimation. The limitation of this system lies in the fact that vehicle-based system cannot provide friction coefficient estimation if the slip-ratio and slip angle are very small, so the signal to noise ratio is not adequate [33] for reliable measurement. There are limitations of vehicle-dynamics-based estimation methods because they demand either sufficient tire slip or adequate tire slip angle to provide real time updates of the estimates. If the slip ratio and slip angle are extremely small, as it can happen during vehicle coasting on a straight road, the system cannot provide friction coefficient estimate updates [34].

(II) *Wheel shaft-based estimation system model:* provides the friction coefficient estimation using the sensors on an extra wheel mounted on the vehicle. Extra tire on redundant wheel can skid and therefore friction values can be estimated. This system has complicated set-ups [33].

TABLE I
Road Surface Condition Estimation Techniques Comparison

| Techniques | Sensor | Signal Processing | Accuracy | Cost |
|---|---|---|---|---|
| Imaging [39,40,41] | Camera | Yes | >90% | High |
| Imaging [39,42] | Optical Sensor/ Laser | Yes | 98% | High |
| Slip-slope [39] | Accelerometer, pressure sensor | Yes | >90% | High |
| Tire Vibration [43] | Accelerometer | Yes | >90% | Medium |
| Solar Radiation [40] | Infrared, temp. sensor, humidity | Yes | >90% | Medium |
| BP NN[39] | No add. sensor | No | 80-90% | Low |
| Acoustic Method[35,36] | IC Voice recorder | Yes | >99% | Low |

(III) *Tire-based estimation system model:* uses tire deflection (vertical, lateral and longitudinal) measurements by piezoelectric strips embedded in the regular tires of the vehicle [33].

To compare available estimation and measurement techniques, a comprehensive list is shown in Table I. The acoustic method shown in Table I uses only IC voice recorder to collect direct information from the road. It is a simple, low cost and fast method. The recognition accuracy can reach to 99%.

In this part, we propose an alternative acoustic based road friction estimation method demonstrating almost all the advantages of the method mentioned in [35, 36]. In addition to that, ARTE system proposed in this work is integrated as a part of traction control systems (TCS) and quantitatively proven to improve the performance of all three TCS structures examined here. More details on ARTE system could be found in [37, 38] and the focus of this work is limited to the integration of ARTE with traction control systems and its performance.

ARTE system is designed as a low-cost solution to deal with the uncertainty in the estimation process of road-tire interaction. The friction force here is estimated to involve the mu-lambda relationship, representing the effective friction coefficient versus the slip ratio. The estimation is performed by recognition of the road type and then finding the corresponding mu-lambda (i.e. friction coefficient vs slip-ratio) look-up tables. To detail more, in ARTE set-up, a cardioid microphone is used to collect the acoustic data from tire/road interaction and the data is processed to classify the road type. This information is then used to select the pre-recorded correct mu-lambda to produce a better optimal lambda value than the assumption in original model which usually features a road-tire model with constant lambda. When the real/actual values do not match the assumed constant values, the performance of the system is undermined. Eliminating this blind assumption and providing a reliable estimation on the actual road-friction, ARTE leverages the system performance. In other words, ARTE updates the model to get closer to the actual dynamics by removing some part of the uncertainty related to the road-type and conditions. The benefit of ARTE can be also explained from an uncertain system perspective. By integrating ARTE, uncertainty bounds of the system become narrower allowing the model to control the traction force much better with a more robust structure. Here, the ARTE model consists of four parts: (a) data collection set-up (b) pre-processing, noise cancellation and signal conditioning, (c) extraction and selection of features, (d) road type estimation using ANN or SVM. The flow-chart in Fig 1 illustrates the signal flow and the processing steps to reach the optimum lambda.

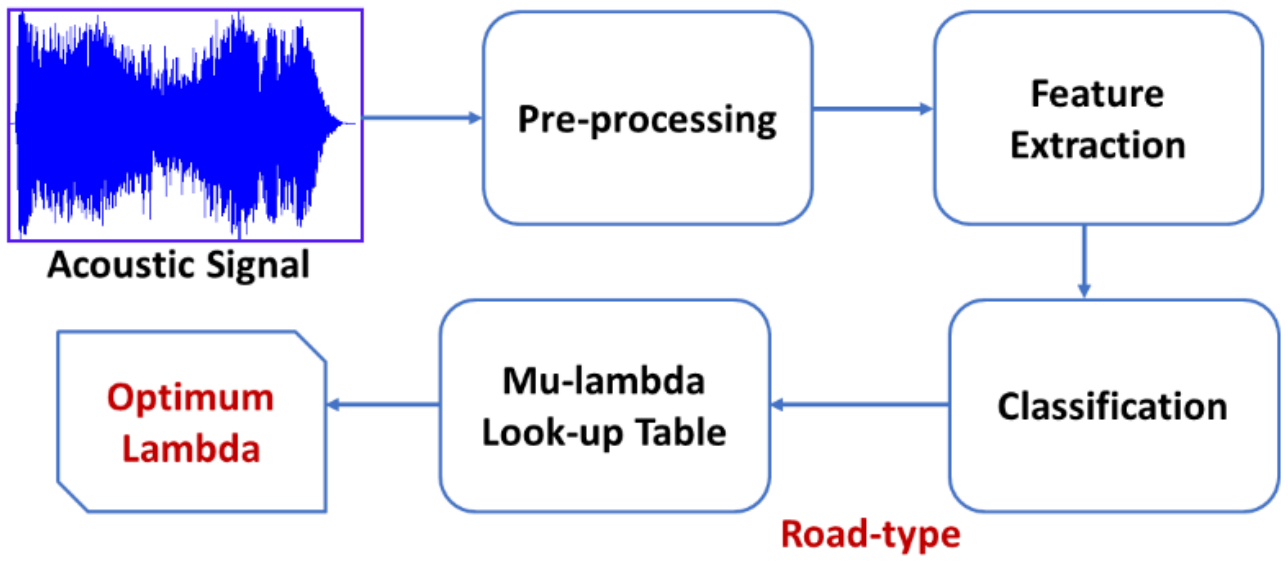

Fig. 1. Flow-chart for operation principle of ARTE unit

## A. *Data Collection Set-up and Pre-processing*

Acoustic data was collected by using a full Electric Vehicle located at Mechatronics Education and Research Centre, ITU, a DC-AC converter, a DPA 4012 model cardioid microphone, a generic microphone amplifier, and a laptop which has Pentium(R) Dual-Core CPU T4200 2GHz processor. Gold Wave program was employed for data acquisition and pre-processing. Acoustic data of gravel, asphalt, snow and stone road types collected in the set-up are shown in Fig 2. The electric vehicle is more suitable than a vehicle with internal combustion engine for ARTE method. Since EVs provide quieter environment, no problems were met regarding the motor noise or combustion chamber. In addition to this, a foam rubber shield was used to suppress the wind noise that could be generated when the vehicle moves. Fig 2 also shows cardioid microphone pattern which graphically demonstrates microphone's directionality. This pattern indicates that the microphone collects data mostly at its front part but to a less extent on its sides. The acoustic data was picked up at constant and low-speeds from 10-30 km/h during collection. Generally, the tire-road friction estimation structures need variable and high-speed profile for the estimation process, however, ARTE system does not demand variable or high-speed profile. In fact, it can be used in all speed profile conditions, including constant and variable speed values.

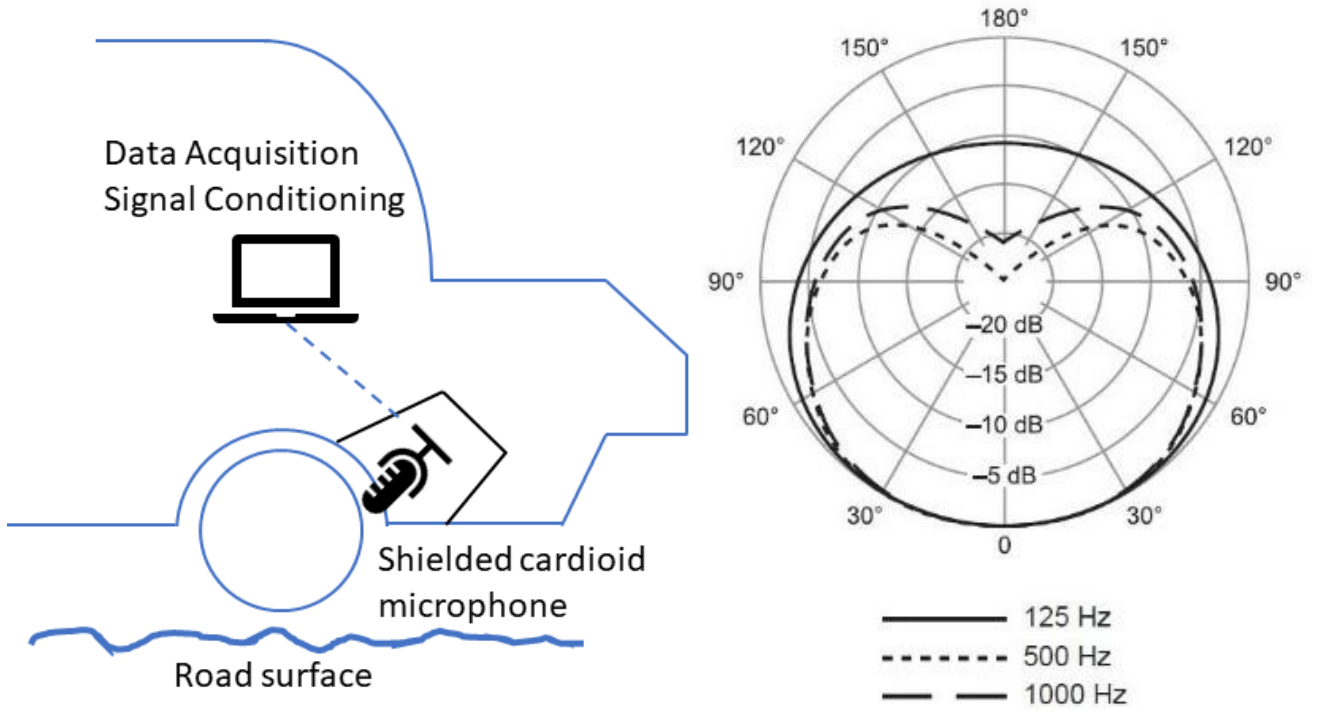

Fig. 2. Data collection and microphone set-up behind EV (left), Example cardioid response (right)

## B. *Pre-processing, Noise Elimination and Signal Conditioning*

The original data is collected under ideal conditions in a test track at ITU Campus without any external noises such as car pass, tire squeal or rain drop noises. However, to test the effect of extra external noises on the acoustic signal recognition performance, the original data with two most common noise sources have been mixed with the noise data provided by [44], some example noises can be seen in Figure 3.

Filtering experiments are performed to see if the noise data can be eliminated to recover back the original data without any significant alterations to the signal characteristics. To achieve this, fundamental frequencies of original and noise data were calculated. Then, by a simple band-pass or low-pass filtering, the noise data could be eliminated. Furthermore, the worst case is also explored to see if the noise was not eliminated whether the classification of road type data would be still possible.

Acoustic data could be classified correctly even though velocity was increased. The data clusters representing different road conditions are still distinguishable; however, a new set of classifiers would be trained for more accurate results if very high speeds are in question [42].

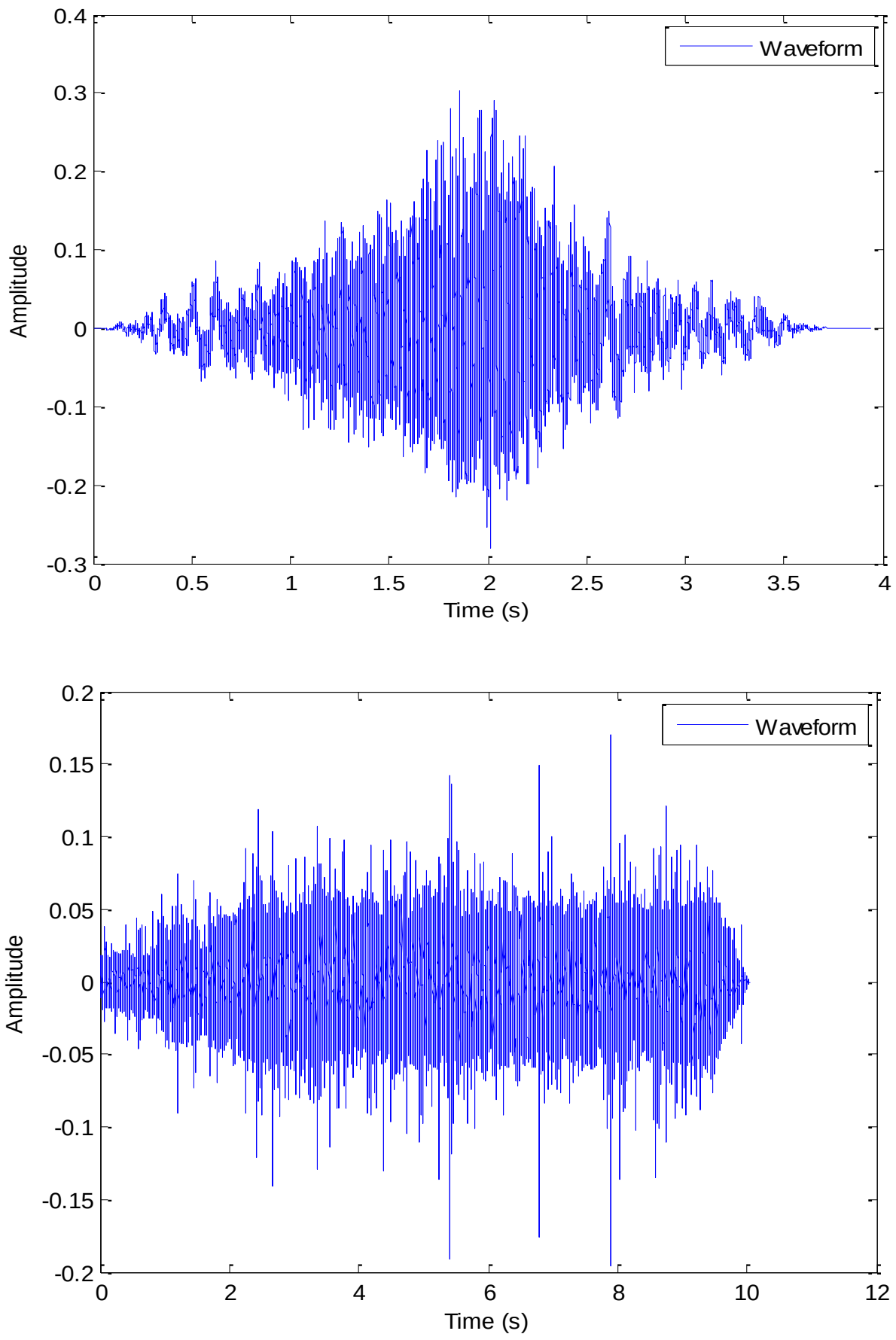

Fig. 3. Sample noise data representing car passing-by (top), rain (bottom)

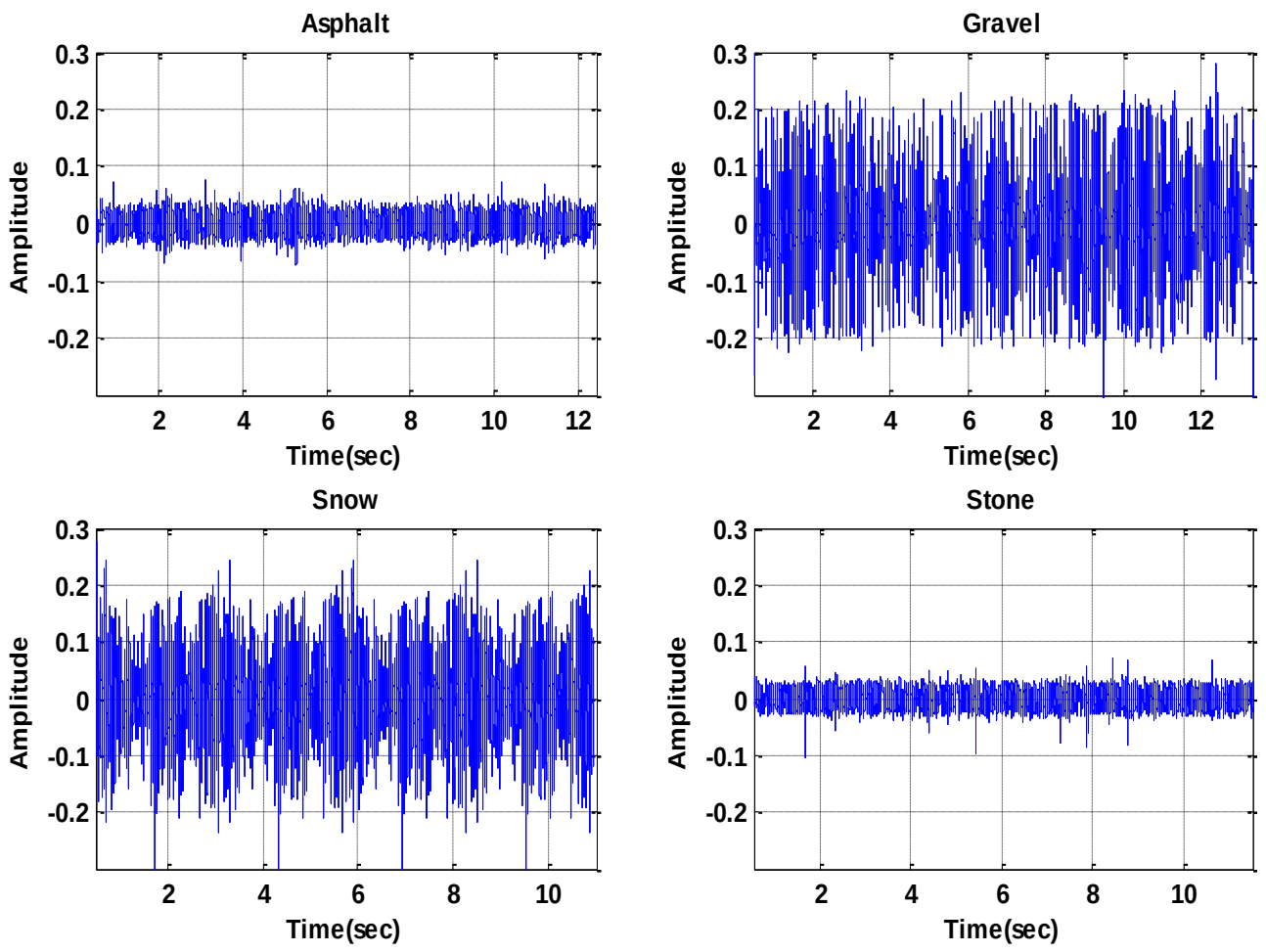

Fig. 4. Raw acoustic signals of asphalt, snowy road, stony road and gravel obtained by ARTE

## C. *Feature Extraction and Selection*

In Fig 4, a sample from the original data of acoustic signals which are acquired by the cardioid microphone of ARTE system can be seen in raw-format data of gravel, asphalt, snow and stony road. Thirty samples were taken for each data record with 0.1 seconds intervals randomly. A feature vector was formed using power spectrums, *cepstrums* and linear predictive coding coefficients (LPC) as acoustic signal processing techniques. These techniques provided to determine the best feature vector which represents audio data. Selection criteria of feature vector elements are minimum variance intra-class coherence and maximum distance criteria for inter-class separability. Firstly, 10 LPC coefficients, 5 power spectrum coefficients and 5 *cepstrum* coefficients are selected to form a feature vector having 20 elements. Then, the feature vector was trimmed using a least variance approach considering intra-class values. After this pruning process, the features providing a coherent class representation remained as 3 LPC, 2 power spectra co-efficient and 2 *cepstrums* values giving a feature vector of seven elements per data point.

In addition to this feature selection, Kullback-Leibler distance was calculated for determining the distance between different road data clusters to observe the effectiveness of the measurement and separability. If the stony road data is taken as the benchmark, the distance between the stony road and asphalt is 936.58, stony road and snowy road is 971.88 and stony road and gravel is 928.13. These Kullback-Leibler distances indicate that the clusters are separable after the most significant digit. After this rough analysis for separability of feature vectors on inter-class and intra-class coherence, ANN and SVM are trained for classification which is detailed in next section.

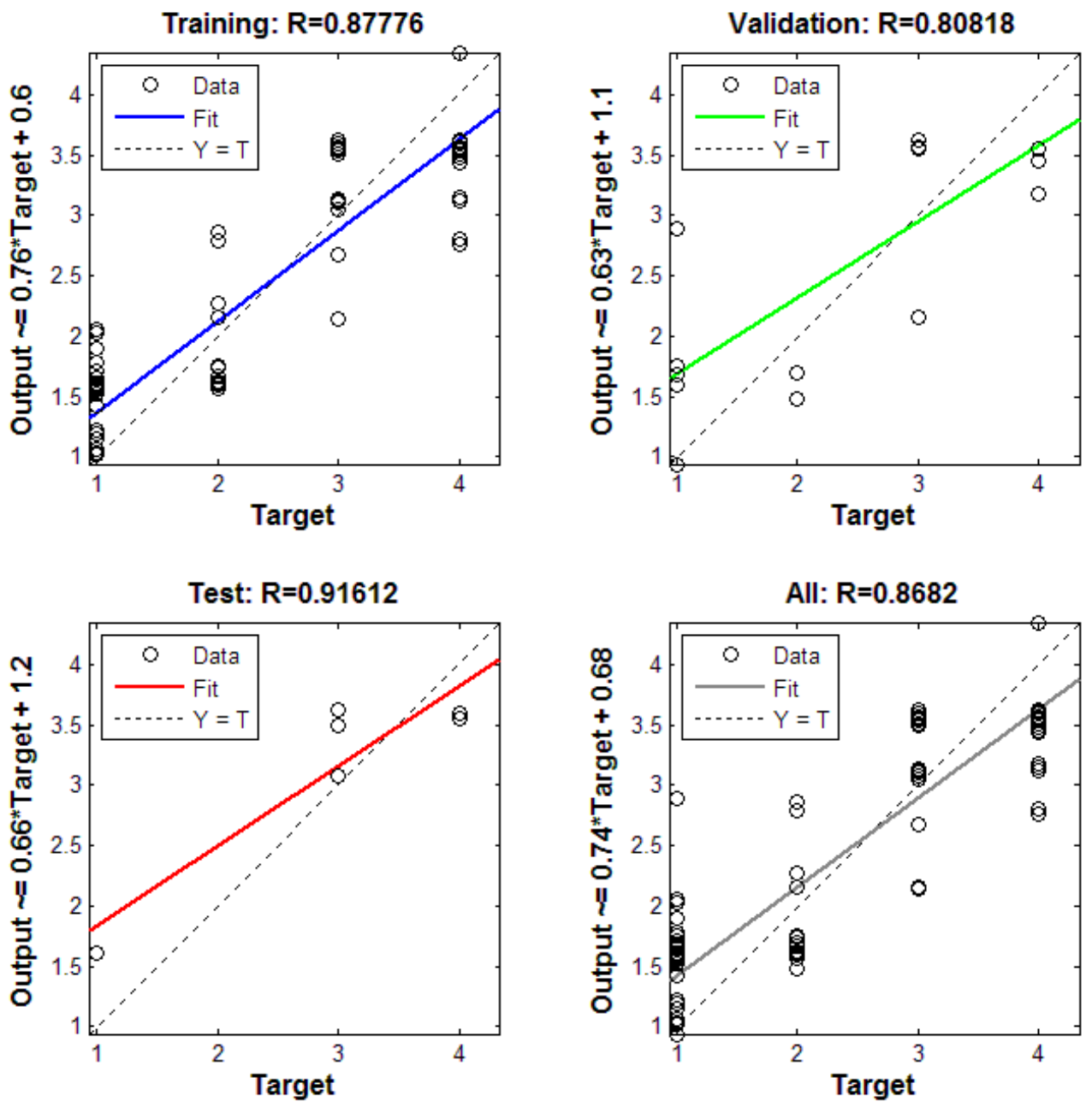

Fig. 5. Regression results of ANN classification, the recognition rate goes up to 91% in test data

### D. *Road Type Estimation using ANN and SVM*

An ANN was trained and tested off-line on the audio data using the selected feature vector. The ANN uses Levenberg-Marquardt learning method and has 4 layers with hidden layers including [4-3-2] neurons in order. The structure is a MLP network and back propagation is used to feedback the errors at each epoch. After 18 epochs, the ANN converged to performance criteria of the mean square error. The regression results in Fig 5 show that ANN can identify the test data with a regression coefficient of 0.91. The coefficient can be increased if the bias in feature vector space is removed. In fact, the results shown here is one of the worst, better performances were obtained giving regression coefficients above 0.95, however that required further pruning and analysis on the vector space.

TABLE II
CONFUSION MATRIX OF THE ANN CLASSIFIER

| ANN Classification | | Actual: Asphalt | Snow | Stone | Gravel | *False Positive* |
|---|---|---|---|---|---|---|
| Predicted | Asphalt | **28** | 0 | 1 | 1 | ***2*** |
| | Snow | 0 | **26** | 3 | 4 | ***7*** |
| | Stone | 1 | 3 | **26** | 2 | ***6*** |
| | Gravel | 1 | 1 | 0 | **23** | ***2*** |
| ***False Negative*** | | ***2*** | ***4*** | ***4*** | ***7*** | |

The performance of the classification can also be summarized in the confusion matrix given in Table II. From this table, classifier has relatively reasonable number of true positive and true negative results out of 30 samples. The correct classification rate is around 85%. Although the system may need improvement, ARTE system would provide a reasonable optimum slip ratio at most of the time if this ANN is included as pattern classifier. It should be noticed that the ARTE method uses the results from the ANN classifier of the road types and matches them with pre-loaded/pre-recorded mu-lambda curves for the optimal lambda estimation (i.e. where the correspondent friction coefficient is the highest) for each road condition. Therefore, a direct estimation of road adhesion coefficient is not performed, rather the road type is identified by the classification. The classification gives the road type, and the slip ratio can be estimated from velocity measurements if they are available in the given TCS. Then, using these two pieces of information (i.e. the road type and slip ratio) the correct mu-lambda look-up table can be used to read the correspondent *mu* (i.e. friction coefficient). In case the system does not have the sensors to obtain the chassis and the wheel velocity, the slip ratio cannot be obtained. The approach is then to take the peak mu value of the corresponding road condition for our reference. Although it seems to introduce a big error by assuming the peak mu-value for all the slip ratios, it still gives better results compared to the traction control systems without road type estimators. The reason is that the road-type classification at least provides the information on the actual road condition and guides the system to take the correct mu-lambda table as the reference. In other words, the error it would have by not selecting the right slip-ratio/ mu couple within a certain road type data is much less than the error it would have if the wrong mu-lambda look-up table had been used.

To explore the possibility of a better classification result, we have trained support vector machines (SVM) on the same data. A well-known benefit of SVM is the unique and global solution while ANN can suffer from multiple local minima. In addition to this, SMVs have the advantage of having a simple geometric interpretation and giving a sparse solution. Unlike ANN, SVM computational complexity does not depend on the input space dimensionality. SVMs use structural risk minimization, while ANNs use empirical risk minimization. The SVMs often outperform ANNs in practice since they are less prone to overfitting [45].

During training, SVM is used to separate three dry road surface condition (gravel, stone, asphalt) from snow using *cepstrums* in a smaller data size. *Cepstrums* are generally more discriminative features than others (i.e. LPC, Δ) because their inter-class variance is the highest. Using SVM classification, road types can be separated well. However, gravel, stone and asphalt data can be occasionally confused. This less successful result of SVM classification can be explained due to using of smaller data and limited features, i.e. *cepstrum*.

A confusion matrix for SVM classification test is given in Table III to compare against the results of ANN in Table II. As it can be seen from Table III, SVM classifier is better than ANN classifier in giving no false negatives while yielding 100% true positives. Despite this successful result, the SVM classifier can also give false positives. However, the frequency of the mistake from false positives is acceptable for road type estimation since the estimation is continuously performed. Therefore, any momentary misclassification will be corrected in a prompt manner in the long-run. It should be also noted that, the false positives can be explained by the operation principle of this algorithm as SVM classification operates on single support vectors which may not always have clear-cut separations between classes all throughout the feature space, therefore, yielding more false positives than expected. For example, when original data from asphalt, stone, gravel and snow was presented to support vector decision surfaces, the classes can be perfectly separated. However, when data from other classes are presented to a single support vector for a particular class, it may falter and conclude in false positives due to imperfections in projections to hyper-space.

TABLE III
CONFUSION MATRIX OF THE SVM CLASSIFIER

| SVM CLASSIFICATION | | **ACTUAL** | | | | |
|---|---|---|---|---|---|---|
| | | ASPHALT | SNOW | STONE | GRAVEL | ***FALSE POSITIVE*** |
| **PREDICTED** | ASPHALT | **20** | 0 | 3 | 2 | ***5*** |
| | SNOW | 0 | **20** | 1 | 1 | ***2*** |
| | STONE | 0 | 0 | **20** | 2 | ***4*** |
| | GRAVEL | 0 | 0 | 6 | **20** | ***6*** |
| ***FALSE NEGATIVE*** | | ***0*** | ***0*** | ***0*** | ***0*** | |

Although both ANN and SVM based road-type estimation algorithms give acceptably good response, they need to be tested for the abrupt change in road conditions to ensure a stable transient response. For this purpose, a road profile having abrupt changes between stone-gravel-asphalt was processed. The samples for extracting the features were taken for 0.1 sec time interval. The interval of the sampling directly affects the transient response of the classification system. It was observed that both road type classification algorithms (ANN and SVM) were able to predict the real road condition with above 95% correct classification rate. The abrupt road condition change did not have any visible transient side-effect. The classification algorithm was able to catch up the real condition within 0.2 sec (i.e. two-folds of the sampling interval) at its worst case.

## III. TRACTION CONTROL SYSTEMS FOR ELECTRIC VEHICLES

In this section, TCS structures are analysed in the order of simplest to most-complex: model following controller (MFC), slip ratio controller (SRC) with the fuzzy inference of the optimal slip ratio, and the maximum transmissible torque estimation (MTTE) model which has an observer for actual driving force. The core data analytics of any TCS consists of tire/road friction force estimation and control. However, tire/road friction forces cannot be directly estimated or controlled. Thus, observers or indirect estimation systems are used in each TCS model. Some of these methods allow better results as stability and robustness (i.e. MTTE and SRC), while the other TCS may have a simpler but rough model (i.e. MFC). Although friction value is not possible to measure directly, some models make it possible to estimate the friction force between the road-tire tire using more available and explicit but empirical tire friction models such as Pacejka and LuGre [46]. Herein, first the principles of MFC, SRC and MTTE traction control systems are introduced with block diagrams. Then, the transfer functions including the torque or the current as input and velocity of the wheel, chassis, and the slip ratio as the output is derived. Next, the performance of these three TCS structures under constant torque demand condition is evaluated. Additional stability conditions are also derived to compare the systems in terms of their dependency on the controller and/or vehicle dynamics parameters. This analysis also helped to see the limitations of the examined TCS model structures independent of the ARTE unit and how much the inclusion of ARTE could improve them.

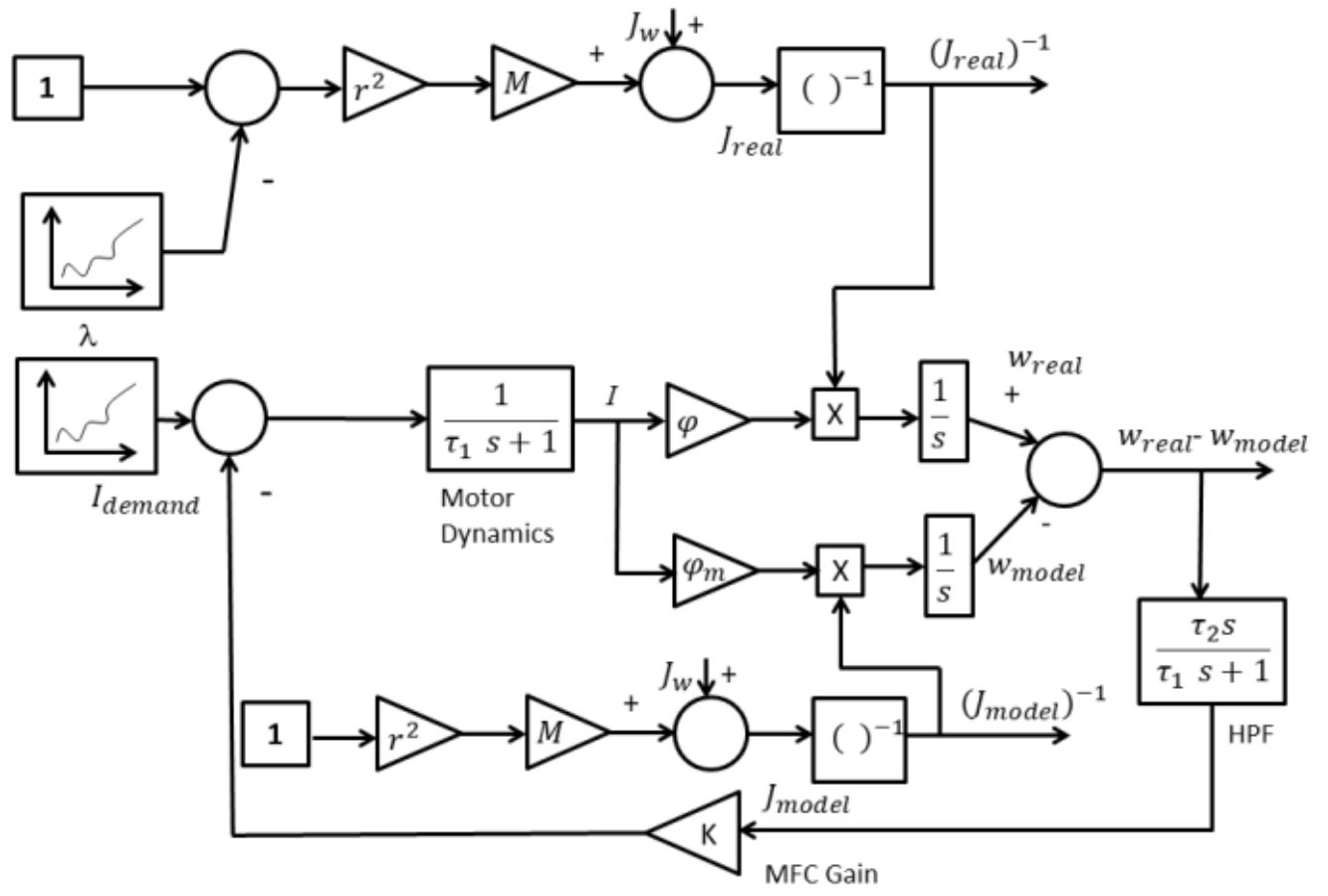

Fig. 6. The detailed Simulink block-diagram of MFC for analysis, based on original model described in [1]

## A. Model Following Controller (MFC)

The aim of MFC model is to make the dynamics of real vehicle follow the structure output which is described as ideal. However, when the real result and model result are different and far away from each other, the actuators might be overloaded and even saturated. The detailed block diagram of MFC based on [1] can be seen in Fig 6.

From the detailed block diagram in Fig 6, it can be understood that the real value of the moment of inertia reflected in the wheels deviate from the ideal value in the model because of the multiplication of the $Mr^2$ term with *(1-λ)*. This means that the real vehicle is felt lighter on the wheels causing the deviation from ideal model. In fact, the requirement on the slip-ratio being zero is a very conservative and rough approach. Any information on the optimal slip ratio for the road will help obtaining a better tracking performance and less load on the actuators. In a classical MFC application, there are two kinds of uncertainty: (i) modeling errors, (ii) measurement noise in the angular velocity. To show the dependency of performance on these two uncertainty sources, transfer functions of the MFC are derived having the torque (or current) and measurement noise as input and angular velocity of the wheel as output.

We start with basic formulae for real and model values of the moment of inertia for a wheel in (1) and (2).

$$J_{real} = J_w + M.r^2(1-\lambda) \quad (1)$$

$$J_{model} = J_w + M.r^2 \quad (2)$$

Then using the detailed block-diagram given in [7], the transfer function between the reference current for motor and the angular velocity difference is given in (3).

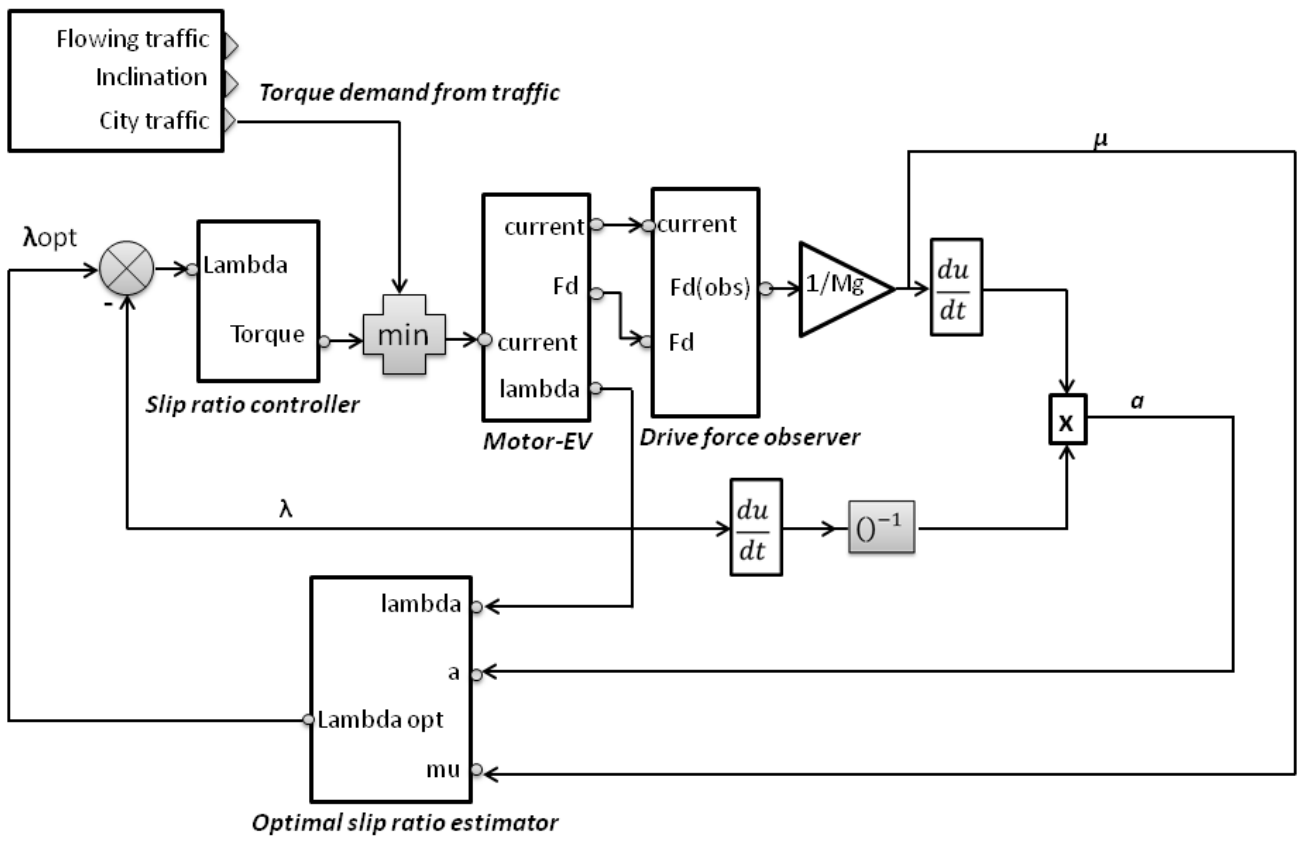

Fig. 7. Slip Ratio Controller structure in open forms

$$\frac{\omega_{diff}(s)}{I_{ref}(s)} = \frac{K.(J_m - J_r)}{s[\tau_1.\tau_2.J_m.J_r.s^2 + (J_m.J_r).(\tau_1+\tau_2).s + (J_m.J_r) + K_{MFC}.\tau_2.K.(J_m - J_r)]} \quad (3)$$

Using the characteristic equation of the transfer function in (3), absolute stability condition in (4) can be derived.

$$K_{MFC} \leq \frac{J_m(\tau_2-\tau_1)^2}{4.\tau_2^2.\tau_1.K.\Delta} \quad (4)$$

The term "Δ" is the uncertainty between the real and the model values of the vehicle inertia described by equation (5).

$$J_r = \frac{J_m}{1+\Delta} \quad (5)$$

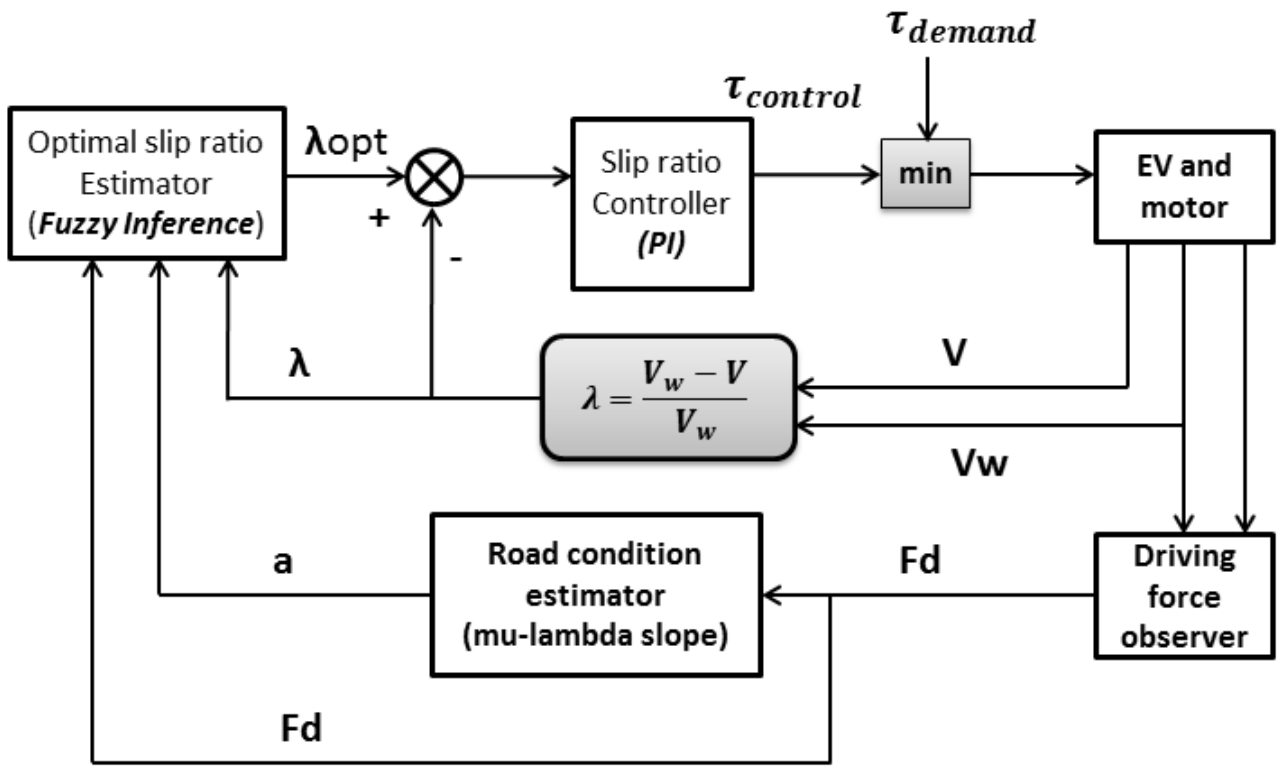

Fig. 8. Detailed SRC system with main parts in block-diagram format

As it can be seen, the MFC gain is an upper bounded variable and it depends on the vehicle inertia $J_m$, motor time constant $\tau_1$, motor gain $K$, HPF time constant $\tau_2$ and most importantly the uncertainty in the vehicle inertia denoted by Δ. The classical MFC uses a predefined $K_{MFC}$ gain value which may not be the optimal value for the situation.

## B. Slip-ratio Controller (SRC)

Slip ratio controller structures in literature has two versions:

(I) SRC with road condition estimator using a free wheel to estimate mu-coefficient and determine the optimum lambda accordingly.

(II) SRC with fuzzy system to estimate the optimum lambda value using the mu-lambda curve slope *a*, *mu* and lambda at that point.

In the analysis here, the second type of SRC with fuzzy system is examined and the optimal slip ratio stands for the value where the correspondent friction coefficient takes maximum attainable value. The closed block diagram of SRC as it appears in [10] and corresponding Simulink diagram in simplified form is given in Figure 7.

As it can be seen from Fig 8, the system uses a kind of optimal slip ratio estimator (i.e. fuzzy inference) and tries controlling the slip-ratio directly by computing and applying the appropriate torque input. Since the slip ratio is more directly addressed in SRC and dynamically included with the estimation, it is a finer method than MFC. In Figure 4, SRC system main control structure can be examined in more detail showing five main parts: (i) EV-motor for the plant to be controlled, (ii) driving observer, (iii) road condition estimator, (iv) slip ratio estimator and (v) controller having PI action.

The related transfer function and its parameters are shown in equations (6) - (8).

$$G(s) = \frac{\lambda(s)}{\tau(s)} = \frac{1-\lambda_o}{a.N_e}\frac{1}{1+\tau s} \quad (6)$$

$$\tau = \frac{M_w V_w}{a.N_e} \quad (7)$$

$$N_e = \frac{M_w + M(1-\lambda_o)}{M}.N \quad (8)$$

Using these relations, a transfer function between the motor current and velocity difference is obtained in (9) with real inertia defined in (10.

$$\frac{V_{diff}(s)}{I(s)} = \frac{r.\lambda}{s.(\tau_m s+1)[Mr^2(1-\lambda)+J_r]} \quad (9)$$

$$J_{real} = \frac{J_n}{1+\Delta}. \quad (10)$$

## C. Maximum Transmissible Torque Estimation (MTTE)

MTTE model does not demand the vehicle chassis velocity or wheel velocity measurements. Observed driving force is the main part of the model as seen in block diagram in Fig 9. First, MTTE model has an observer to obtain the driving force $F_d$. Then, the system uses this observed driving force to calculate and obtain the maximum transmissible and allowable torque. One of the advantages of this model is that MTTE uses the acceleration ratio of the vehicle chassis and wheel as a relaxation value and tries to keep the ratio as close as possible to one or greater. It means that since the vehicle chassis and wheels accelerations are not allowed to deviate much, the speeds would not be so different from each other as well, thus guaranteeing a low slip ratio.

To examine the robust stability of the MTTE model, in [12] an equivalent closed-loop block-diagram is given (see *Appendix 2* for details) with an additive uncertainty in the inertia of the vehicle together with derived $\boldsymbol{T_{zw}}$ transfer function (11). However, the derivation of the transfer function is not detailed here for the sake of brevity. Further examination of transfer function $\boldsymbol{\tau(s)/\tau_{max}(s)}$ can be performed, relating the relaxation variable alpha, $\boldsymbol{\alpha}$, and the slip ratio value, lambda, $\boldsymbol{\lambda}$ using the full model in [12].

TABLE IV
UNITS FOR MAGNETIC PROPERTIES

| Parameter | Magnitude | Units |
|---|---|---|
| ***Jw:*** moment of inertia for the Wheel | 0.6 | $kg.m^2$ |
| ***r:*** wheel Radius | 0.28 | m |
| $\boldsymbol{\tau_1}$: Time constant for electric motor | 0.05 | sec |
| $\boldsymbol{\tau_2}$: Time constant for high pass filter in MFC | 1000 | sec |
| ***M:*** *vehicle mass* | 1400 | kg |
| ***m:*** *wheel mass* | 10 | kg |

$$T_{zw} = \frac{-J_w.K}{J_n r\tau_1\tau s^2 + J_n(r\tau - K\tau_1 + r\tau_1)s + J_n r - Mr^2K} \qquad (11)$$

## IV. Performance Analysis of TCS Structures with ARTE

The TCS suggested previously [1-4, 7] are simulated under several dangerous scenarios, i.e. slippery road conditions and frequent change of road characteristics which may cause instability. Vehicle parameters given in Table IV are used for simulating all TCS structures to establish a common comparison ground representing the same vehicle.

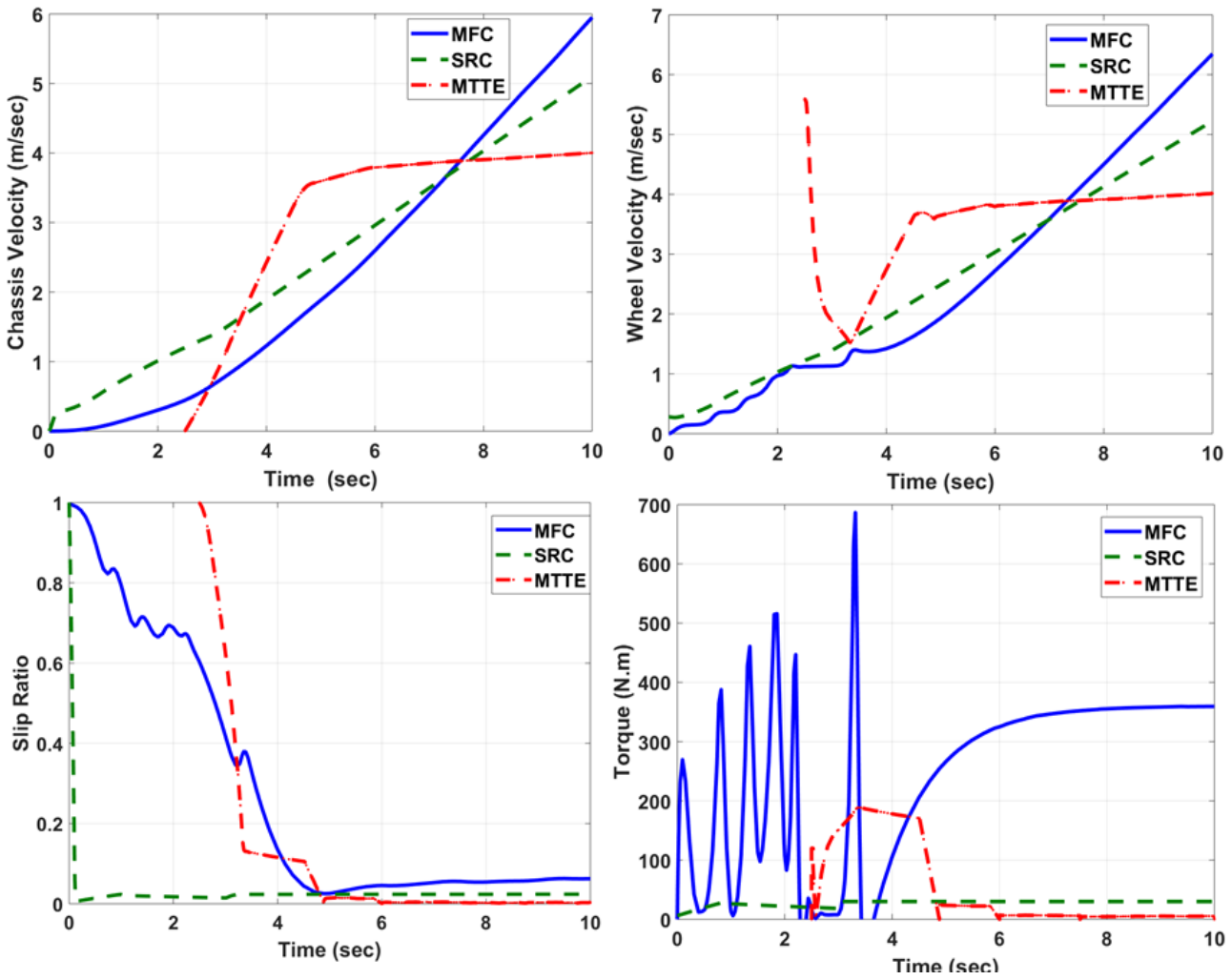

Fig. 10. MFC, SRC and MTTE traction control systems ***without ARTE***

To compare these three TCS in a systematically reasonable manner, we have selected four criteria indicating the performance of a TCS: (i) time averaged total deviation of slip ratio from zero (i.e. the ideal value), (ii) the max torque of the motor during TCS in action, (iii) the normalised area under the torque-time graph showing the correction effort and energy consumption, (iv) gap metric to assess the stability under worst cases of deviation from the nominal system parameters. The study considered only constant torque demand profiles herein, however, several other traffic conditions involving inclination and stop-and-go traffic conditions can be also simulated using the models. It must be noted that the models in this work consider only linear regions in traction control problem. The application is implemented in a bracket of 10-30 km/h speed under limited test track requirements. Performances of MFC, SRC and MTTE systems without using ARTE module are shown in Figure 10.

ARTE system with a trained and tuned classifier is integrated in all TCS structures in the scope of this work. The integration was performed for three TCS structures as follows: For MFC, the total inertia of the vehicle used in the model is updated according to the estimated value of lambda in $J=Jw+Mr^2(1-\lambda)$. In the SRC structure, the integration is straightforward since the estimated optimum lambda by ARTE is directly needed in the algorithm as the reference signal. Lastly, the MTTE system uses the estimated result in updating the $\boldsymbol{\alpha}$ coefficient used to calculate the maximum torque.

After the integration, evaluation was performed in a series of simulated tests. The simulations mainly consist of road scenarios where different mu-lambda characteristics and dynamic torque-demands may occur. The ARTE module is integrated in MFC to update the reference model, whereas in SRC as the reference optimal slip ratio source and in MTTE to update the relaxation factor. The main mechanism behind why ARTE helps improving the TCS performance is that it reduces the uncertainty sources or narrows down the uncertainty bounds in the systems. The comparison of three TCS structures after the integration with ARTE can be seen in Fig 11. The results are only presented for the constant torque demand, however the other road conditions demanding different torque profiles have similar results, therefore not included here for brevity.

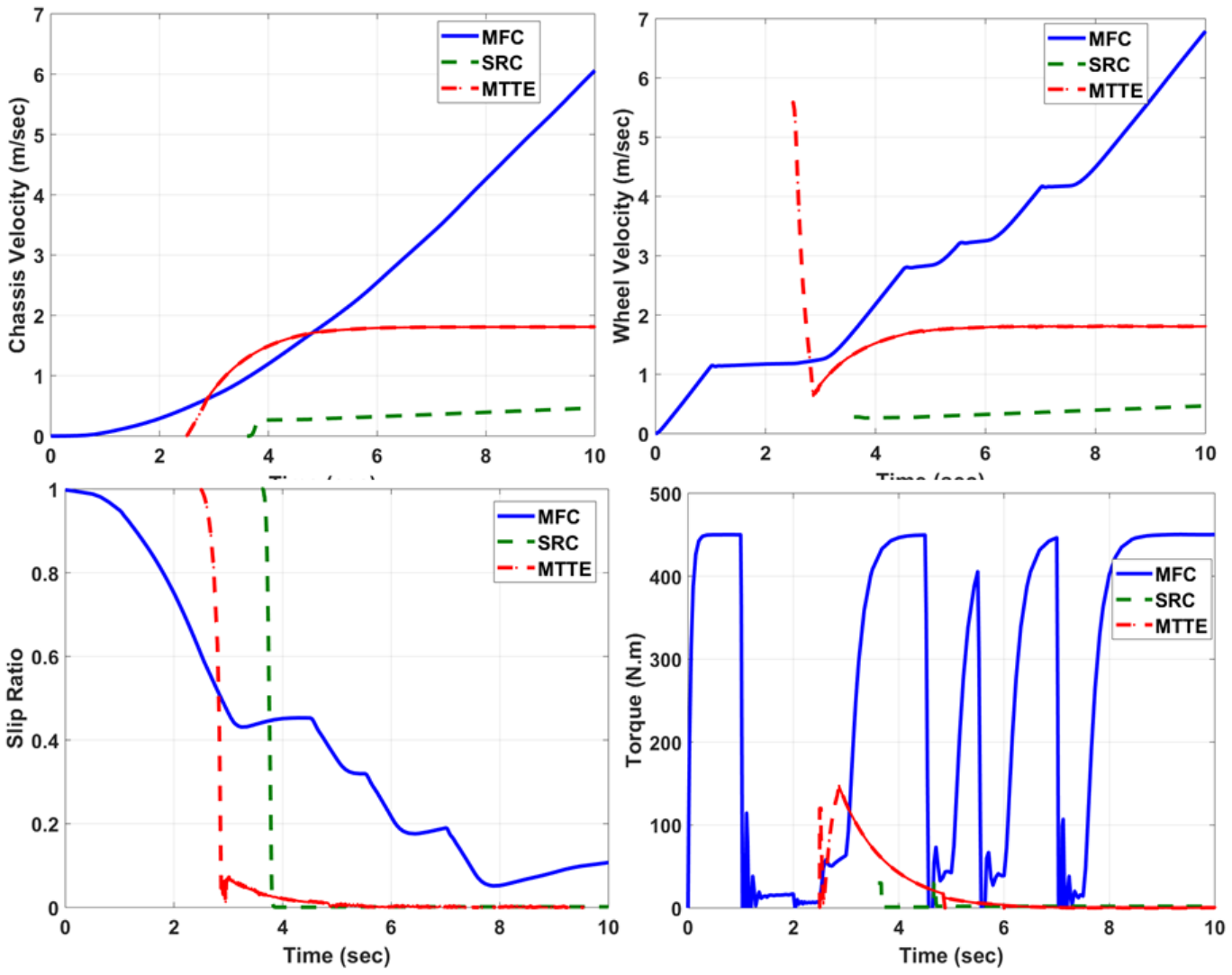

Fig. 11. MFC, SRC and MTTE traction control systems ***with ARTE***

Fig. 9. MTTE block-diagram in closed form

TABLE V
PERFORMANCE COMPARISON OF TCS WITH AND WITHOUT ARTE

| Traffic-Road Condition | TCS Structure | Time-averaged slip ratio deviation | Max torque during operation | Normalized area of torque-time graph |
|---|---|---|---|---|
| **Constant torque demand** | **MFC** | 0.4224 | 686.80 | 212.30 |
| | **SRC** | 0.0654 | 300.00 | 258.21 |
| | **MTTE** | 0.1278 | 188.58 | 43.10 |
| | **MFC-*ARTE*** | 0.4672 | 450.29 | 173.48 |
| | **SRC-*ARTE*** | 0.0237 | 300.00 | 33.44 |
| | **MTTE-*ARTE*** | 0.0522 | 147.11 | 26.83 |

The performance criteria measured on all TCS structures with and without ARTE module is given in Table V. As it can be seen from the table and the Figures 10-11, the slip ratio is reduced to a level very close to zero especially when SRC and MTTE is integrated with ARTE. It was also observed that although ARTE helped MFC system to have less angular velocity difference between the model and the real vehicle, the slip ratio reflected a smaller improvement. The SRC system also benefited from ARTE; however, the average and max torque values were still high compared to MTTE. As a result, it can be concluded that MTTE with ARTE is the best TCS structure of all since the slip ratio has reduced without a formidable increase in the torque correction effort.

It should also be noted that MTTE structure with/without ARTE has a delay in reacting to environment changes (Fig.10-11) as implementation of ARTE in SRC structure introduced a similar delay as can be noticed in Fig.11.

In addition to performance criteria results given in Table V, *gapmetric* defined by Vinnicombe [47] is also measured between the nominal and uncertain plant models in MFC, SRC and MTTE structures. It was seen that SRC was more prone to instability since the gap between the nominal and uncertain plant can go up to 0.91 in SRC structure, whereas it remains relatively small in MTTE with 0.85 and the smallest value is calculated for MFC structure being just 0.0332. This analysis perfectly matches the characteristics of three structures since MFC is crude but less prone to uncertainties since no estimation of lambda or dynamical value is included in the structure. The SRC structure gives the worst performance in terms of *gapmetric* because it contains a double estimation process with more uncertain parameters. MTTE lies between these two extremes; however, it also suffers from the uncertainty in observation process. After the integration with ARTE system, the robustness of all systems has greatly improved, reducing the gap metric from 0.0332 to 0.0127 for MFC; from 0.91 to 0.28 for SRC and from 0.85 to 0.096 for MTTE systems.

## V. CONCLUSIONS

As an active vehicle safety system, the application of traction control systems (TCS) for electric vehicles (EV) is feasible due to advantages of the torque control with direct-drive motors. However, TCS structures usually require the estimated road-tire friction and slip-ratio as input. Obtaining the road-tire friction might require some intricate model-based indirect methods and/or expensive sensors. Furthermore, the systems using the slip ratio explicitly as input or feedback require the measurement of accurate chassis and wheel velocity. Another roadblock in front of effective TCS application is that existing TCS structures are often designed without any consideration of the robustness and energy efficiency of torque control. In this study, we have considered only three TCS structures for EVs and improved their performances with the integration of ARTE. Out of these systems, only the SRC structure requires measurement of chassis and wheel velocity since it explicitly uses the slip ratio as input. MFC and MTTE on the other hand does not need the chassis velocity. We addressed the robustness and energy-efficiency issues while reducing the uncertainty in internal models or assumptions in the control structures. The system features a low-cost road-type estimation unit is based on acoustic signals picked up from the vicinity of the interface between the road and the tire of the EV. The design of this acoustic road-type estimator (ARTE) unit is presented including data collection set-up, feature extraction, signal processing and classification modules. To demonstrate the effect of ARTE on TCS structures, it is integrated into three existing TCS models for supplying the uncertainty bounds in the system for the optimum slip ratio. Next, the performances of these three TCS models are evaluated using the same physical parameters of the vehicle for constant torque demand scenario. As a result, it was found that based on the simulations, ARTE system can improve the performances of SRC and MTTE models greatly while it does not necessarily contribute towards the performance of MFC structures. The system is able to reduce the slip-ratios up to 75% in general.